\documentclass[twocolumn,floatfix, nofootinbib,prd,preprintnumbers,showpacs,showkeys,superscriptaddress,longbibliography]{revtex4-2}
\usepackage{slashed}
\usepackage{amsmath}
\usepackage{booktabs}
\usepackage{array}

\usepackage{tikz}
\usetikzlibrary{positioning}
\usetikzlibrary{calc}
\usepackage{graphicx} 
\usepackage{orcidlink}
\usepackage[top=2cm,bottom=2cm,left=1cm,right=1cm,marginparwidth=1.75cm]{geometry}
\title{Asy Static Finite Distance}
\usepackage{mathrsfs}
\usepackage{amssymb}
\usepackage{amsmath}
\date{\today}
\usepackage{mathtools}
\usepackage{comment}

\usepackage[mathscr]{euscript}
\usepackage{amsmath}
\usepackage{graphicx}
\usepackage{dcolumn}
\usepackage{bm}
\usepackage{epsfig}
\usepackage{amssymb,latexsym,mathrsfs}
\usepackage{graphicx}
\usepackage{color}
\usepackage{hyperref}
\usepackage{diagbox}

\usepackage{xcolor}
\definecolor{darkblue}{RGB}{0,0,150}

\usepackage{textgreek}

\usepackage{tikz}

\usepackage{amsmath}
\usepackage{physics}
\usepackage{csquotes} 

\hypersetup{
    colorlinks=true,
    linkcolor=red,
    citecolor=blue,
}

\usepackage{amsmath}
\usepackage{amssymb}
\usepackage{subfigure}
\usepackage{hyperref}
\usepackage{url}
\usepackage{xcolor}
\usepackage{color}
\usepackage{soul}
\definecolor{amaranth}{rgb}{0.9, 0.17, 0.31}
\definecolor{purple(munsell)}{rgb}{0.62, 0.0, 0.77}
\definecolor{americanrose}{rgb}{1.0, 0.01, 0.24}
\definecolor{palatinateblue}{rgb}{0.15, 0.23, 0.89}
\definecolor{royalblue(web)}{rgb}{0.25, 0.41, 0.88}
\definecolor{hanpurple}{rgb}{0.32, 0.09, 0.98}
\definecolor{beaublue}{rgb}{0.74, 0.83, 0.9}
\definecolor{carminered}{rgb}{1.0, 0.0, 0.22}
\definecolor{brightpink}{rgb}{1.0, 0.0, 0.5}
\definecolor{vividviolet}{rgb}{0.62, 0.0, 1.0}
\hypersetup{ linktoc=all,
    colorlinks, linkcolor={palatinateblue},
    citecolor={brightpink}, urlcolor={amaranth}}

\usepackage{comment}
\newcommand{\Ei}{\operatorname{Ei}}
\newcommand{\be}{\begin{equation}}
\newcommand{\ee}{\end{equation}}
\newcommand{\bs}{\begin{split}} 
\newcommand{\bea}{\begin{eqnarray}}
\newcommand{\eea}{\end{eqnarray}}

\newcommand{\bes}{\begin{subequations}}
\newcommand{\ees}{\end{subequations}}

\newcommand{\bo}{\raise-1mm\hbox{\Large$\Box$}}

\usepackage{booktabs}
\usepackage[table]{xcolor}
\definecolor{lightgray}{gray}{0.9}
\definecolor{lightblue}{rgb}{0.8,0.85,1}

\begin{document}


\title{Self-Reflection in a Moving Mirror}

\author{Michael R.R. Good\, \orcidlink{0000-0002-0460-1941}}
\email{muon@asu.edu}
\affiliation{Physics Department \& Energetic Cosmos Laboratory, Nazarbayev University,\\
Astana 010000, Qazaqstan.}
\affiliation{Leung Center for Cosmology \& Particle Astrophysics,
National Taiwan University,\\ Taipei 10617, Taiwan.}
\affiliation{Beyond Center for Fundamental Concepts in Science, Arizona State University,\\
Tempe AZ 85287, USA.}

\author{Eric V.\ Linder\,\orcidlink{0000-0001-5536-9241}  }
\email{evlinder@lbl.gov} 
\affiliation{Berkeley Center for Cosmological Physics \& Berkeley Lab, University of California,\\ Berkeley CA 94720, USA.}

\begin{abstract} 
 We present an analytic flat-spacetime accelerating boundary analog of Hawking-type emission that 
 possesses infinite asymptotic acceleration (and radial acceleration in the black hole analog) but finite total  radiated energy (and zero surface gravity of the black hole). We perform 
a unified study of its scattering symmetry, horizon formation, asymptotically extreme acceleration, finite total radiated energy, and the distinction between local energy flux and global particle production within a single closed-form model. The particle spectrum, energy spectrum, and equivalent spacetime metric are derived, revealing an interesting mix of normal and extremal black hole properties.

\end{abstract}
\maketitle
\tableofcontents


\section{Introduction}

\[
\begin{tikzpicture}[
    scale=0.9,
    every node/.style={font=\normalsize},
    dot/.style={circle,fill=black,inner sep=2.2pt},
    box/.style={draw, rounded corners=4pt, inner sep=4pt, align=center, fill=white}
]

\coordinate (E) at (0,2.8);
\coordinate (T) at (6,2.8);
\coordinate (N) at (3,0);

\draw[line width=0.9pt]
    (E) -- node[midway, above=2pt] {} (T)
        -- node[midway, sloped, above=2pt] {$\alpha\to\textrm{const.}$} (N)
        -- node[midway, sloped, above=2pt] {$E \to \mathrm{const.}$} (E);

\node[dot,label={[box]above:Thermal black holes\\ $(\alpha,E) \to \infty$}] at (E) {};
\node[dot,label={[box]above:Extremal black holes\\ $(\alpha, E) \to \textrm{const.}$}] at (T) {};
\node[dot,label={[box]below:$(\alpha, E) \to (\infty,\textrm{const.})$\\ This note.}] at (N) {};

\end{tikzpicture}
\]

Acceleration is a central quantity in physics, with 
Einstein's Equivalence Principle teaching us that it can 
be mapped equivalently to gravity or spacetime curvature. 
Accelerating, nongravitating systems such as moving mirrors \cite{DeWitt:1975ys,Davies:1974th,Davies:1976hi,Davies:1977yv,Ford:1982ct,carlitz1987reflections,CW2lifetime,Ford:1990ae,Holzhey:1994we}
have thus been fruitfully employed to elucidate relations 
concerning black holes and spacetime horizons.

Moving mirrors offer a simple but versatile framework for probing a range of nontrivial problems. Among other applications, they have been used to study the information-loss problem~\cite{Hotta:2015yla,PisinChen2017,Reyes:2021npy,Agullo:2025opy}, entanglement harvesting~\cite{akal2021entanglement,kawabata2021probing,osawa2024final,wald2019particle,Cong:2020nec,Cong:2018vqx}, fluctuation--dissipation~\cite{jackel1992fluctuations,Hsiang:2024xlh,Xie:2023wvu}, holography~\cite{akal2021holographic,akal2022zoo,ievlev2024moving,ling2022reflected,pujolas2008strongly,sato2022complexity}, and conformal field theory~\cite{biswas2024moving,Biswas:2024mlq,Kumar:2023kse}. More broadly, particle creation induced when moving boundaries couple to the quantum vacuum, known as the dynamical Casimir effect (DCE)~\cite{moore1970quantum}, has generated a substantial body of work; see~\cite{dodonov2009dynamical,dodonov2010current,Dodonov:2025rxz} for reviews. This broader literature likewise supports the utility of various boundary-types of moving-mirror models~\cite{dodonov1996generation,alves2003dynamical,alves2006dynamical,alves2008energy,alves2010exact,good2021quantum}, including studies of partially reflecting mirrors~\cite{barton1993quantum,obadia2001notes,nicolaevici2001quantum,haro2008black,nicolaevici2009semitransparency,fosco2017dynamical,Lin:2021bpe}, and effects like quantum radiation-reaction forces~\cite{jaekel1993quantum,barton1995quantum1,barton1995quantum2,lambrecht1996motion,jaekel1997movement,alves2008quantum,Alves:2009ev,butera2019mechanical,Gorban:2024vss}, as well as experimental verification~\cite{wilson2011observation,lahteenmaki2013dynamical,vezzoli2019optical,paraoanu2020listening}, such as beta decay \cite{Lynch:2022rqx,Ievlev:2023inj,Good:2025qta}, plasma mirrors \cite{Chen:2020sir,Chen:2015bcg} and the AnaBHEL collaboration \cite{AnaBHEL:2022sri,Navick:2024wbd,Hsiung:2025mya,Chen:2025xkv}.

These developments have clear physical signatures in quantum particle production. The accelerating-boundary correspondence, see e.g., \cite{Good:2020byh,Good:2020uff,GoodMPLA,Good:2021dkh,Lin:2024ihr,Good:2019tnf,Mujtaba:2024vqy,Mujtaba:2024vmf}:
\[
\text{mirror acceleration} \;\to\; \text{spacetime horizon} \;\to\; \text{black hole}
\]
provides an effective framework for relating trajectory dynamics to energy flux, including recent understanding of negative-energy flux \cite{Bianchi:2026xoi}, and to particle production. 

For example, thermal (Hawking) radiation from black holes \cite{Hawking:1974rvNATURE,Hawking:1974sw,Wald:1975kc,Page:1976df}
can be directly mapped \cite{wilczek1993quantum} to certain asymptotically infinite 
acceleration systems \cite{Good:2016oey}, while extremal black holes involve 
acceleration asymptotically approaching a constant value \cite{Liberati_2000,Liberati:2000zz,Rothman:2000mm}. 
In the former case, the total radiated energy is infinite, while in the latter case, it is finite. Here, we investigate 
whether the connection between infinite acceleration and 
infinite energy can be broken, and what sort of black hole this implies. 
(We know the converse is true: an asymptotically constant 
acceleration will lead to a cessation of radiation and 
does not give infinite total energy \cite{good2020extreme}.)

We study relativistic exponentially accelerated motion (see the extremal-like or `super-remnant' analog trajectory, Eq.~20 of \cite{Good:2018aer}) in which the proper acceleration equals the rapidity, yielding an asymptotically infinite acceleration, to see whether it can yield finite particle energy. This model  possesses three remarkable features:

\begin{itemize}
    \item Involution symmetry: the light-cone coordinate map is invariant under the exchange of advanced and retarded time.
    \item Finite horizon evaporation: a perfect reflector along this motion approaches a null-horizon with infinite acceleration and emits a finite amount of radiation to an observer at future null infinity. 
    \item Consistent stress-quanta: the total energy carried by the particles agrees with that computed from the stress tensor.
\end{itemize}

For simplicity, we use the dimensionless acceleration scale, $\kappa = 1$, and natural units where $G = \hbar = c = 1$.

\section{Elements of our Involution Trajectory}


\subsection{Proper-time coordinates} 

We explore a moving mirror model with asymptotic infinite acceleration but finite energy, with the kinematic property
\be
\text{acceleration}=\text{rapidity},
\tag{2}
\ee
or, more explicitly,
\be
\alpha(\tau)\equiv \frac{d\eta}{d\tau}=\eta(\tau),
\ee
where $\eta(\tau)$ is the signed rapidity. Defining the rapidity magnitude by
\be
\rho(\tau)\equiv |\eta(\tau)|,
\ee
and choosing the left-moving branch (by the usual convention, see e.g., \cite{Fabbri}), means $\eta<0$.  Thus, we have
\be
\frac{d\eta}{\eta(\tau)}= d\tau,
\tag{3} \qquad \rightarrow\qquad \eta(\tau)=-e^{\tau},
\ee
after setting the acceleration scale to unity. Equivalently,
\be
\rho(\tau)=e^{\tau},
\qquad
|\alpha(\tau)|=\rho(\tau).
\tag{4}
\ee
As we will show using light cone coordinates, this simple exponentially accelerated motion possesses an interesting and unexpected scattering symmetry.
\subsection{Space-time coordinates}
\begin{figure}[h] 
    \centering     \includegraphics[width=0.40\textwidth]{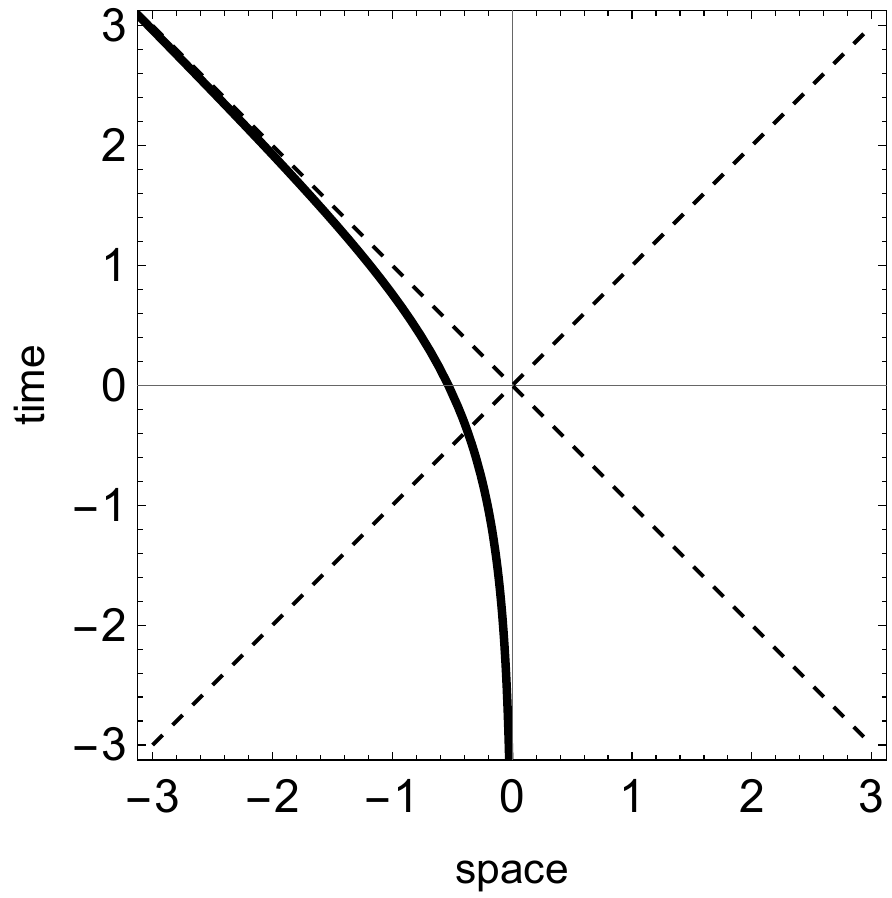}
    \caption{Space-time trajectory of the mirror, Eq.~(\ref{spacetimeofrho}), in \((z,t)\) coordinates, with unit scale, plotted parametrically as a function of the rapidity magnitude \(\rho>0\) running along the curve. The horizon has been set by convention to the advanced time \(v=0\), corresponding to the left-moving null line through the origin. The mirror is taken to move to the left, giving the standard appearance of any ordinary thermal black-hole trajectory. In the asymptotic past, it is at rest at the spatial origin, \(z=0\), and then accelerates toward the horizon. The dashed lines are the null rays \(t=\pm z\). There is no visual hint in the causal structure that the energy emission would be finite.} \label{fig:spacetime} 
\end{figure}

Starting from the proper-time description, the rapidity determines the
velocity through
\begin{equation}
\textrm{v}(\tau) = \tanh \eta(\tau).
\end{equation}
The space-time coordinates follow from
\begin{equation}
\frac{dt}{d\tau} = \cosh\eta(\tau), \qquad
\frac{dz}{d\tau} = \sinh\eta(\tau).
\end{equation}
Using the rapidity $\eta(\tau) = -e^{\tau}$, the trajectory is obtained by
integrating
\begin{equation}
t(\tau) = \int \cosh(e^{\tau})\, d\tau, \qquad
z(\tau) = -\int \sinh(e^{\tau})\, d\tau .
\end{equation} 
Since $d\tau = d\rho/\rho$, the integrals can be written
\begin{equation}
t(\rho) = \int \frac{\cosh\rho}{\rho}\, d\rho, \qquad
z(\rho) = -\int \frac{\sinh\rho}{\rho}\, d\rho .\label{spacetimeofrho}
\end{equation} 
The trajectory is shown in a spacetime diagram in 
Fig.~\ref{fig:spacetime}. 
These expressions naturally lead to representations in terms of
exponential–integral functions.  It is therefore convenient to express the trajectory in null coordinates, which we introduce next.

\subsection{Light-cone coordinates}

In null coordinates,
\be
u=t-z,
\qquad
v=t+z,
\ee
the trajectory may be written parametrically in terms of the rapidity magnitude \(\rho>0\) as
\be
u(\rho)=\operatorname{Ei}(\rho),
\qquad
v(\rho)=\operatorname{Ei}(-\rho).
\ee
Accordingly, the ray-tracing functions are
\be
p(u)=v,
\qquad
f(v)=u.
\ee
Since \(f\) and \(p\) are inverse maps, one has $p=f^{-1}$. But interestingly, this light-cone trajectory is its own self-inverse. Indeed, eliminating \(\rho\) gives
\be
p(u)=\operatorname{Ei}\!\left(-\operatorname{Ei}^{-1}(u)\right),
\qquad
f(v)=\operatorname{Ei}\!\left(-\operatorname{Ei}^{-1}(v)\right),
\ee
and therefore
\be
p(p(u))=u,
\qquad
f(f(v))=v.
\ee
Thus, the scattering map is an involution: applying it twice returns the original null coordinate. In other words, the trajectory exhibits a null-coordinate exchange symmetry between advanced and retarded times, \(u\leftrightarrow v\).

\subsection{Scattering symmetry}

For a moving mirror in $1+1$ dimensions, the light-cone coordinate ray-tracing function $p(u)=v$ relates retarded time $u = t-x$ to advanced time $v=t+x$. For a mirror at rest, this map is an involution, $p(p(u))=u$; or scattering symmetry such that
\be\textrm{retarded times}\; u \quad \longleftrightarrow \quad  \textrm{advanced times}\; v.\ee
This property is most apparent for the static mirror, where $p(u) = u$ and $f(v) = v$, which is the most physically trivial example of such a symmetry, producing neither redshifted field modes nor particles. And of course, a mirror moving at constant velocity can be Lorentz transformed into a static mirror and produces no particles (though it is not formally involutive).  A centered hyperbola of uniform acceleration, $u v = -1$, is involutive, but produces no energy. 

It is therefore of interest that our exponential trajectory preserves 
the same involutive scattering structure while still allowing acceleration to generate particle-energy creation.

\subsection{Stress-tensor energy flux}

The energy flux at future null infinity is determined by the Schwarzian derivative, see e.g, \cite{Birrell:1982ix}
\begin{equation}
F(u) = -\frac{1}{24\pi}\{p(u),u\}.
\end{equation}
Using the trajectory $p(u)$, one has a flux that can be written
\begin{equation}
F(u) = \frac{\rho e^{-2\rho}}{12\pi},
\end{equation}
where $u = \operatorname{Ei}(\rho)$. The flux rises from zero, reaches a peak, and then decays exponentially. The total energy emitted by the mirror is
\begin{equation}
E = \int_{-\infty}^{\infty} F(u)\,du .
\end{equation}
Using
\begin{equation}
du = \frac{e^\rho}{\rho} d\rho,
\end{equation}
one obtains
\begin{equation}
E = \frac{1}{12\pi}\int_0^\infty e^{-\rho} d\rho = \frac{1}{12\pi}. \label{stress-tot-energy}
\end{equation}
Thus, the mirror emits a finite amount of energy while accelerating in an unlimited manner.

\subsection{Double-log-scaling of proper time}

One other characteristic of interest especially for the black hole analog is that, at late retarded times, the growth of proper time grows only double-logarithmically with $u$.  
Starting from the null-coordinate relation
\begin{equation}
u(\rho)=\operatorname{Ei}(\rho), \qquad \rho=e^{\tau},
\end{equation}
we use the large-$\rho$ asymptotic form of the exponential integral,
\begin{equation}
\operatorname{Ei}(\rho)\sim \frac{e^{\rho}}{\rho}, \qquad \rho\to\infty.
\end{equation}
Thus, for late times,
\begin{equation}
u \sim \frac{e^{\rho}}{\rho}
\quad \Longrightarrow \quad
\ln u \sim \rho-\ln\rho \sim \rho.
\end{equation}
Therefore, since
\begin{equation}
\rho \sim \ln u, \quad \textrm{and}, \quad \rho=e^{\tau},
\end{equation}
we obtain, as $u\to\infty$,
\be \tau \sim \ln\ln u. \label{double-log}\ee
This shows that the mirror’s proper time grows exceptionally slowly relative to retarded time. By itself, this is a kinematic property of the trajectory, but it suggests a possible analogy with the long-throat structure of extremal horizons, where the proper distance to the horizon diverges. We defer the geometric interpretation until Section~\ref{sec:metric_correspondence} when the metric correspondence is explicitly established.

\section{Particle Emission}
\subsection{Bogoliubov coefficients}

Particle production is encoded in the Bogoliubov coefficients between
incoming and outgoing modes.  Starting from the standard moving-mirror
expression in terms of the ray-tracing function \(p(u)\), see e.g., \cite{Good:2016atu}
\begin{equation}
\beta_{\omega\omega'}
=
\frac{1}{4\pi\sqrt{\omega\omega'}}
\int_{-\infty}^{\infty}
du\,
e^{-i\omega u-i\omega' p(u)}
\left(\omega' p'(u)-\omega\right),
\end{equation}
with \(\omega>0\) and \(\omega'>0\).  Integration by parts gives
\be
\beta_{\omega\omega'}
=
-\frac{1}{2\pi}
\sqrt{\frac{\omega}{\omega'}}
\int_{-\infty}^{\infty}
du\,
e^{-i\omega u-i\omega' p(u)},
\ee
where the boundary term is dropped. Putting our trajectory into this integral expressed in terms of the rapidity magnitude \(\rho\), where
\begin{equation}
u=\Ei(\rho), \qquad p(u)=\Ei(-\rho), \qquad \rho>0\ ,
\end{equation}
gives
\be 
\beta_{\omega\omega'}
=
-\frac{1}{2\pi}
\sqrt{\frac{\omega}{\omega'}}
\int_{0}^{\infty}
\frac{e^\rho}{\rho}\,
\exp\!\left[-i\omega \mathrm{Ei}(\rho)-i\omega' \mathrm{Ei}(-\rho)\right]
d\rho.
\ee
This integral cannot be easily solved but can be put  
into a simpler form that points to some key properties. 
Introducing
\begin{equation}
\nu=\sqrt{\omega\omega'},
\qquad
\lambda=\frac12\ln\!\left(\frac{\omega'}{\omega}\right),
\end{equation}
so that \(\omega=\nu e^{-\lambda}\) and \(\omega'=\nu e^{\lambda}\), the
Bogoliubov coefficient becomes
\be
\beta_{\nu\lambda}
=
-\frac{e^{-\lambda}}{2\pi}
\int_{0}^{\infty}
\frac{e^\rho}{\rho}\,
\exp\!\left[
-i\nu\,S_{\lambda}(\rho)
\right]
d\rho,
\ee
where the phase is
\begin{equation}
S_{\lambda}(\rho)
=
e^{-\lambda}\Ei(\rho)
+
e^{\lambda}\Ei(-\rho).
\end{equation}

Differentiating the phase yields 
\be
S'_\lambda(\rho)
=
e^{-\lambda}\frac{e^\rho}{\rho}
+
e^\lambda\frac{e^{-\rho}}{\rho}
=
\frac{2\cosh(\rho-\lambda)}{\rho}. \label{eq:sprime} 
\ee
The integral prefactor may be written in purely hyperbolic form as
\be
e^{-\lambda}\frac{e^\rho}{\rho}
=
\frac{1}{2}\bigl(1+\tanh(\rho-\lambda)\bigr)\,S'_\lambda(\rho).
\ee
Substituting this into $\beta_{\nu\lambda}$ gives
\be
\beta_{\nu\lambda}
=
-\frac{1}{4\pi}
\int_0^\infty
\bigl(1+\tanh(\rho-\lambda)\bigr)
S'_\lambda(\rho)\,
e^{-i\nu S_\lambda(\rho)}
\,d\rho.
\ee
Using the chain rule, 
\be
\frac{d}{d\rho}e^{-i\nu S_\lambda(\rho)}
=
-i\nu S'_\lambda(\rho)e^{-i\nu S_\lambda(\rho)},
\ee
so that
\be
\beta_{\nu\lambda}
=
-\frac{i}{4\pi\nu}
\int_0^\infty
\bigl(1+\tanh(\rho-\lambda)\bigr)
\frac{d}{d\rho}e^{-i\nu S_\lambda(\rho)}
\,d\rho, 
\ee
then we can integrate again by parts.  Discarding the boundary term gives 
\be
\beta_{\nu\lambda}
=
\frac{i}{4\pi\nu}
\int_0^\infty
\frac{d}{d\rho}\bigl(1+\tanh(\rho-\lambda)\bigr)
e^{-i\nu S_\lambda(\rho)}
\,d\rho.
\ee
Since
\be
\frac{d}{d\rho}\bigl(1+\tanh(\rho-\lambda)\bigr)
=
\sech^2(\rho-\lambda),
\ee
one arrives at our ultimate result 
\be
\beta_{\nu\lambda}
=
\frac{i}{4\pi\nu}
\int_0^\infty
\sech^2(\rho-\lambda)\,
e^{-i\nu S_\lambda(\rho)}
\,d\rho. \label{betaRHOLAMBDA}
\ee
This form of the Bogoliubov coefficient will help us understand energy emission from a particle perspective.

\begin{widetext}
\subsection{Quantum summation of the energy}

The total energy should be the sum of the energy of each of the particles, see e.g., \cite{walker1985particle}.  With an energy $\omega$ for each particle, the total energy is written as
\begin{equation}
E
=
\int_0^\infty
\int_0^\infty
\omega
|\beta_{\omega\omega'}|^2
d\omega d\omega'.
\end{equation}
Changing to our variables
$\omega = \nu e^{-\lambda}$ and $\omega' = \nu e^{\lambda}$ gives
\begin{equation}
E
=
\int_0^\infty
\int_{-\infty}^{\infty}
2\nu^2 e^{-\lambda}
|\beta(\nu,\lambda)|^2
d\lambda d\nu.
\end{equation}
For the complex conjugate of $\beta_{\nu\lambda}$, we have
\be
\beta_{\nu\lambda}^*
=
-\frac{i}{4\pi\nu}
\int_0^\infty
\sech^2(\rho'-\lambda)\,
e^{\,i\nu S_\lambda(\rho')}
\,d\rho',
\ee
and the square gives
\be
|\beta_{\nu\lambda}|^2
=
\frac{1}{16\pi^2\nu^2}
\int_0^\infty\!\!\int_0^\infty
\sech^2(\rho-\lambda)\sech^2(\rho'-\lambda)\,
e^{-i\nu \Delta S_\lambda(\rho,\rho')}
\,d\rho\,d\rho'.\label{beta2nulambda}
\ee
where $\Delta S_\lambda(\rho,\rho')
=
S_\lambda(\rho)-S_\lambda(\rho')$. Plugging the beta squared into the energy sum formulation gives the energy as
\be
E
=
\frac{1}{8\pi^2}
\int_0^\infty
\int_{-\infty}^{\infty}
e^{-\lambda}
\left(
\int_0^\infty
\int_0^\infty
\sech^2(\rho-\lambda)\sech^2(\rho'-\lambda)\,
e^{-i\nu\Delta S_\lambda(\rho,\rho')}
\,d\rho\,d\rho'
\right)
\,d\lambda\,d\nu.
\ee
We prepare to integrate over $d\nu$ first,
\be
E
=
\frac{1}{8\pi^2}
\int_0^\infty
\int_0^\infty
\int_{-\infty}^{\infty}
e^{-\lambda}\sech^2(\rho-\lambda)\sech^2(\rho'-\lambda)
\left(
\int_0^\infty e^{-i\nu\Delta S_\lambda(\rho,\rho')}\,d\nu
\right)
\,d\lambda\,d\rho\,d\rho'.
\ee
Integrated distributionally, we obtain
\be
E
=
\frac{1}{8\pi^2}
\int_0^\infty
\int_0^\infty
\int_{-\infty}^{\infty}
e^{-\lambda}\sech^2(\rho-\lambda)\sech^2(\rho'-\lambda)
\left[
\pi \delta\!\bigl(\Delta S_\lambda(\rho,\rho')\bigr)
-
i\,\mathcal{P}\!\left(\frac{1}{\Delta S_\lambda(\rho,\rho')}\right)
\right]
\,d\lambda\,d\rho\,d\rho'.
\ee
The principal-value term is antisymmetric under interchange of the two integration variables,
\be
\Delta S_\lambda(\rho',\rho)=-\Delta S_\lambda(\rho,\rho'),
\qquad
\mathcal{P}\!\left(\frac{1}{\Delta S_\lambda(\rho',\rho)}\right)
=
-\mathcal{P}\!\left(\frac{1}{\Delta S_\lambda(\rho,\rho')}\right),
\ee
while the remaining factor, $
e^{-\lambda}\sech^2(\rho-\lambda)\sech^2(\rho'-\lambda)$ is symmetric under \( \rho\leftrightarrow\rho' \). Since the integration region $(\rho,\rho')\in(0,\infty)\times(0,\infty)$ is also invariant under this interchange, the principal-value contribution integrates to zero. Therefore, only the delta-function term remains:
\be
E
=
\frac{1}{8\pi}
\int_0^\infty
\int_0^\infty
\int_{-\infty}^{\infty}
e^{-\lambda}\sech^2(\rho-\lambda)\sech^2(\rho'-\lambda)\,
\delta\!\bigl(\Delta S_\lambda(\rho,\rho')\bigr)
\,d\lambda\,d\rho\,d\rho'.
\ee
Using the delta-function identity in the variable $\rho'$, and noting that for fixed $(\lambda,\rho)$ the phase difference $\Delta S_\lambda(\rho,\rho')$ has the unique root $\rho'=\rho$ because $S'_\lambda(\rho')>0$, we may write
\be
E
=
\frac{1}{8\pi}
\int_0^\infty
\int_{-\infty}^{\infty}
e^{-\lambda}\sech^2(\rho-\lambda)
\left(
\int_0^\infty
\sech^2(\rho'-\lambda)\,
\frac{\delta(\rho'-\rho)}{S'_\lambda(\rho')}
\,d\rho'
\right)
d\lambda\,d\rho.
\ee
Substituting the trigonometric form
$\rho' S'_\lambda(\rho')=2\cosh(\rho'-\lambda)$ from Eq.~(\ref{eq:sprime}), and carrying out the $\rho'$ integral, which localizes at $\rho'=\rho$, gives
\be
E
=
\frac{1}{16\pi}
\int_0^\infty
\int_{-\infty}^{\infty}
e^{-\lambda}\,
\rho\,\sech^5(\rho-\lambda)
\,d\lambda\,d\rho. \label{eq:Einteglamrho} 
\ee
This form can help us better understand the energy carried away by the created particles, as discussed 
in the next subsection. 
\end{widetext}

\subsection{Energy as a function of kinematic rapidity}
Writing $x=\rho-\lambda$ and $d\lambda=-dx$ we obtain an integral over the rapidity magnitude
\be
E
=
\frac{1}{16\pi}
\int_0^\infty
\rho e^{-\rho}
\left(
\int_{-\infty}^{\infty}
e^{x}\sech^5 x\,dx
\right)
\,d\rho.
\ee
Since the integral inside the parentheses is $4/3$, we are left with particle energy as a function of rapidity magnitude $E(\rho)$, and integrated energy $E$, 
\be
E(\rho) = \frac{1}{12\pi} \rho e^{-\rho}, \qquad E =  \frac{1}{12\pi}\int_0^\infty \rho e^{-\rho}\,d\rho= \frac{1}{12\pi},
\ee
in agreement with the stress-tensor result, Eq.~(\ref{stress-tot-energy}). 

\subsection{Energy as a function of frequency rapidity }
In addition to $E(\rho)$ it can also be useful to consider 
the energy as distributed over the rapidity-like
frequency variable 
\begin{equation}
\lambda = \frac12 \ln\frac{\omega'}{\omega}\ .
\end{equation} 
This variable 
represents the logarithmic frequency shift between incoming and outgoing modes. 
Starting from 
Eq.~(\ref{eq:Einteglamrho}) 
we define the energy density in \(\lambda\) by carrying out the \(\rho\)-integration first:
\be
E(\lambda)
=
\frac{e^{-\lambda}}{16\pi}
\int_0^\infty
\rho\,\sech^5(\rho-\lambda)\,d\rho.
\ee
While not especially illuminating, this may be written in closed form as
\be
E(\lambda)
=
\frac{e^{-\lambda}}{384\pi}
\left[
18\,\operatorname{Ti}_2(e^\lambda)
-9\,\sech\lambda
-2\,\sech^3\lambda
\right], \label{E(lambda)}
\ee
where the arctan integral, $\textrm{Ti}_2$, is related to the dilogarithm combination as
\be
\operatorname{Ti}_2(x)
\equiv
\int_0^x \frac{\arctan t}{t}\,dt
=
\frac{1}{2i}\left[\operatorname{Li}_2(ix)-\operatorname{Li}_2(-ix)\right].
\ee
For our real \(\lambda\), the dilogarithm combination is real, so \(E(\lambda)\) is real. 

\begin{figure}[h] 
    \centering     \includegraphics[width=0.45\textwidth]{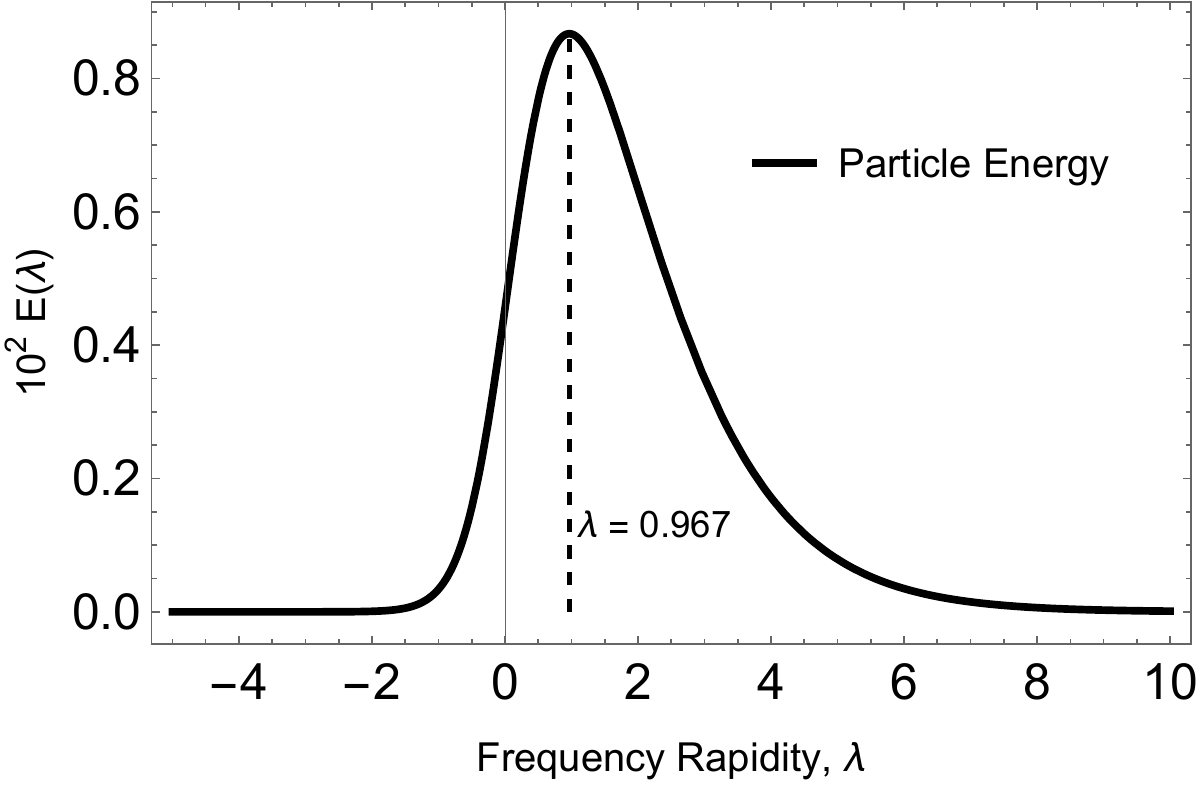}
    \caption{Energy density \(E(\lambda)\) distributed over the frequency rapidity \(\lambda=\frac12\ln(\omega'/\omega)\), from Eq.~(\ref{E(lambda)}). The curve is strongly localized, and note that the maximum occurs at \(\lambda_{\rm max}\approx 0.967154\) (dashed line), showing that the radiated energy is dominated by logarithmic frequency shifts of order one. Integrating the distribution over all \(\lambda\) yields the total energy,
\(\int_{-\infty}^{\infty} E(\lambda)\,d\lambda = 1/(12\pi)\), matching the stress-tensor energy.} 
\label{fig:elam} 
\end{figure}

We show the energy distribution \(E(\lambda)\) in Fig.~\ref{fig:elam}. The rapidity-space energy density is sharply localized and has a unique maximum at $\lambda_{\rm max} \approx 0.967154$, so the peak occurs near, but not exactly at, \(\lambda=1\).  Equivalently,
\be
\frac{\omega'}{\omega}=e^{2\lambda_{\rm max}}\approx 6.927,
\ee
which shows that most of the radiated energy comes from mode pairs whose frequencies differ by about a factor of seven.  The distribution is asymmetric, with tails
\be
E(\lambda)\sim \frac{3}{128}\,\lambda e^{-\lambda}
\quad (\lambda\to+\infty),\ee
and
\be
E(\lambda)\sim \frac{2}{25\pi}\,e^{4\lambda}
\quad (\lambda\to-\infty),
\ee
so negative rapidity shifts are more strongly suppressed than positive ones. 
The total energy is fully analytic, recovered by integrating over all \(\lambda\):
\be
E
=
\int_{-\infty}^{\infty} E(\lambda)\,d\lambda
=
\frac{1}{12\pi},
\ee
also in agreement with the total stress, Eq.~(\ref{stress-tot-energy}).

\subsection{Rapidity interpretation}

The energy kernel,
\be
\sech^5(\rho-\lambda),
\ee
is sharply peaked at
\be
\rho=\lambda,
\ee
and decays rapidly as \(|\rho-\lambda|\) increases. Thus, the dominant contribution to the energy integral comes from the region
\be
\rho \approx \lambda.
\ee
This means that the rapidity magnitude \(\rho\) of the mirror most efficiently couples to the logarithmic frequency ratio \(\lambda\) when the two are approximately equal. In other words, the particle production is largest when the frequency-space rapidity matches the kinematic rapidity of the trajectory, recalling that:
\be
\rho=\frac{1}{2}\ln\!\left(\frac{1+\textrm{v}}{1-\textrm{v}}\right),
\qquad
\lambda=\frac{1}{2}\ln\!\left(\frac{\omega'}{\omega}\right).
\ee
To see the physical content of this matching, notice that
the condition \(\rho\approx\lambda\) is thus equivalent to
\be
e^{2\rho}\approx e^{2\lambda},
\qquad\Longrightarrow\qquad
\frac{1+\textrm{v}}{1-\textrm{v}}\approx \frac{\omega'}{\omega}.
\ee
Here \((1+v)/(1-v)\) is the net Doppler shift associated with reflection from the moving mirror, i.e., the square of the usual one-way relativistic Doppler factor. 
The localization of the kernel around $\rho=\lambda$ gives a direct rapidity interpretation of the quantum summation: the dominant contribution comes from those portions of the trajectory whose kinematic rapidity matches the frequency-space rapidity.

In other words, the mirror radiates most efficiently when, at the level of the integrand factor \(\sech^5(\rho-\lambda)\), the incoming and outgoing frequencies are centered at \(\rho=\lambda\), and therefore particle-energy creation favors configurations in which the kinematic rapidity and frequency rapidity are comparable.  This kind of sharp localization is not generic: for example, thermal mirror radiation \cite{carlitz1987reflections} does not single out such a narrow preferred value of the logarithmic frequency ratio -- 
thermal (Hawking) radiation has constant energy flux, varying rapidity, and a broad distribution of coupling between $\omega$ and $\omega'$. Likewise extremal black holes have no narrow relation between kinematic rapidity and the frequency ratio.

\subsection{Infrared behavior}
For fixed $\lambda$, the Bogoliubov spectrum
(i.e., the particle spectrum, not the energy spectrum) of Eq.~(\ref{beta2nulambda}) in the limit $\nu\to 0$
satisfies
\begin{equation}
|\beta(\nu,\lambda)|^2
\to
\frac{(1+\tanh\lambda)^2}{16\pi^2 \nu^2}.
\end{equation}
This follows because the phase factor approaches unity as $\nu\to 0$, so the leading-order double integral factorizes and evaluates to $(1+\tanh\lambda)^2$. Thus, for fixed $\lambda$, the particle spectrum exhibits a non-integrable $1/\nu^2$ infrared divergence.

This infrared divergence signals an accumulation of soft quanta. For the considered mirror trajectory, the asymptotically unbounded late-time acceleration is physically consistent with the emission of an infinite number of arbitrarily low-frequency particles. This infrared divergence is in particle count rather than in the energy per mode, and does not cause infinite radiated energy.

\subsection{Energy densities in rapidity space}

It is useful to compare the distribution of energy over the rapidity magnitude
\(\rho=e^\tau\) in the stress-tensor and quantum-summation approaches.

\begin{figure}[h] 
    \centering     \includegraphics[width=0.45\textwidth]{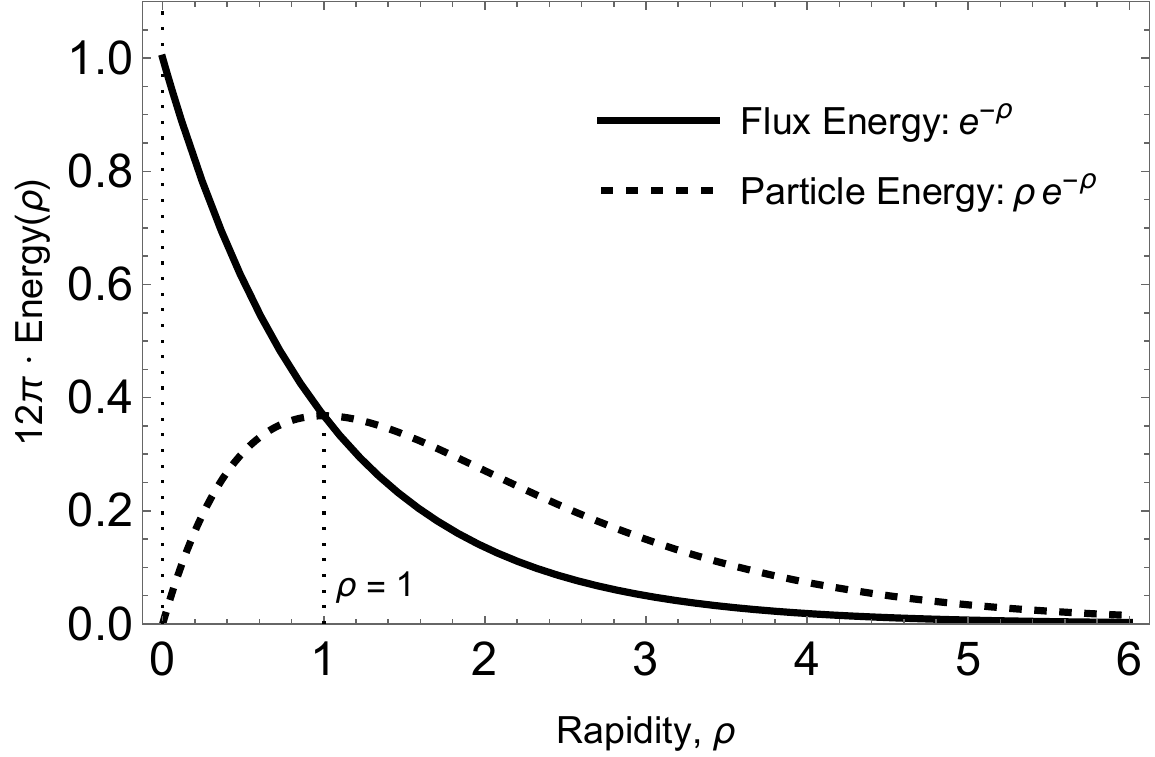}
    \caption{Rapidity-space energy densities \(E_{\mathrm{flux}}(\rho)\) and \(E_{\mathrm{part}}(\rho)\), defined in Eqs.~(\ref{flux}) and (\ref{part}), respectively. Despite these different profiles of rapidity, the area under each curve is identical; both descriptions yield the same total energy, \(E=1/(12\pi)\). 
    \label{fig:energyrho} 
    }
\end{figure}

From the stress-tensor side, the total energy may be written as an integral over
retarded time \(u\),
\be
E=\int_{-\infty}^{\infty} F(u)\,du.
\ee
When we changed variables to \(\rho\), we used $u=\operatorname{Ei}(\rho)$ and $du=(e^\rho/\rho)\,d\rho$ and found the energy density in rapidity space
\be
E_{\mathrm{flux}}(\rho)=\frac{e^{-\rho}}{12\pi},
\qquad
E=\int_0^\infty E_{\mathrm{flux}}(\rho)\,d\rho
=
\frac{1}{12\pi}.\label{flux}
\ee
By contrast, the particles have energy density,
\be
E_{\mathrm{part}}(\rho)=\frac{\rho e^{-\rho}}{12\pi}, \label{part}
\qquad
E=\int_0^\infty E_{\mathrm{part}}(\rho)\,d\rho
=
\frac{1}{12\pi}.
\ee
Both descriptions yield the same total energy, but they distribute that
energy differently over rapidity, 
as illustrated in Fig.~\ref{fig:energyrho}. 
The stress-tensor density
\(\,E_{\mathrm{flux}}(\rho)\propto e^{-\rho}\,\) is largest at \(\rho=0\) and
then decays monotonically, whereas the particle-carrying density
\(\,E_{\mathrm{part}}(\rho)\propto \rho e^{-\rho}\,\) vanishes at \(\rho=0\),
peaks at $\rho=1$ and only then decays exponentially.

The two energy densities provide complementary local and global perspectives on the same radiation process. The global mode decomposition of the particle production introduces an additional rapidity weighting that is absent in the local flux.

\section{Metric Correspondence}
\label{sec:metric_correspondence}
Using the accelerated boundary correspondence (ABC), we identify the null-shell coordinate
with the mirror ray-tracing function,
\begin{equation}
u(\rho)=\Ei(\rho), 
\qquad 
U(\rho)=v(\rho)=\Ei(-\rho).
\label{eq:inv_uU}
\end{equation}
For a null shell at $v=v_0$, the standard matching gives
\begin{equation}
r=\frac{v_0-U}{2},
\qquad
r_*=\frac{v_0-u}{2},
\label{eq:inv_match}
\end{equation}
where $r_*$ is the tortoise coordinate.  The null-shell parameter $v_0$ may be chosen so that the horizon lies at a finite areal radius \(r_H = v_0/2\); in particular, setting \(v_0 = 4M\) places the horizon at \(r_H = 2M\), yielding the standard Schwarzschild horizon area \(A_H = 16\pi M^2\).  Nevertheless, for simplicity, we have chosen $v_0=0$ for the following:
\begin{equation}
r(\rho)=-\frac{1}{2}\Ei(-\rho),
\quad r_*(\rho)=-\frac{1}{2}\Ei(\rho).
\label{eq:inv_r}
\end{equation}
Differentiating these parametric relations,
\begin{equation}
\frac{dr}{d\rho}
=
-\frac{e^{-\rho}}{2\rho},
\quad
\frac{dr_*}{d\rho}
=
-\frac{e^{\rho}}{2\rho},
\label{eq:inv_drstar}
\end{equation}
we can find the tortoise differential,
\begin{equation}
\frac{dr_*}{dr}
=
\frac{dr_*/d\rho}{dr/d\rho}
=
e^{2\rho}.
\label{eq:inv_ratio}
\end{equation}
Considering a static spherically symmetric line element,
\begin{equation}
ds^2=-F(r)\,dt^2+F(r)^{-1}dr^2+r^2d\Omega^2,
\label{eq:inv_static_metric}
\end{equation}
for which the tortoise coordinate satisfies
\begin{equation}
\frac{dr_*}{dr}=F(r)^{-1}.
\label{eq:inv_tortoise_def}
\end{equation}
Comparing Eqs.~\eqref{eq:inv_ratio} and \eqref{eq:inv_tortoise_def}, we obtain
\begin{equation}
F(r)=e^{-2\rho}.
\label{eq:inv_Frho}
\end{equation}
Since Eq.~\eqref{eq:inv_r} implies
\begin{equation}
\rho=-\Ei^{-1}(-2r),
\label{eq:inv_rho_of_r}
\end{equation}
the metric function becomes
\begin{equation}
F(r)=\exp\!\bigl[2\,\Ei^{-1}(-2r)\bigr].
\label{eq:inv_Fr}
\end{equation}
While the exact self-inverse symmetry belongs to the null ray-tracing map, in the curved-spacetime parametrization it survives in the weaker form \(r_*(\rho)=r(-\rho)\).

\subsection{Hovering Acceleration and Surface Gravity}

What can we learn about this new geometry? Let us first find the location of the horizon of this metric, where $F(r_H)=0$.
Using
\begin{equation}
r(\rho)=-\frac{1}{2}\operatorname{Ei}(-\rho),
\end{equation}
we see that as $\rho\to\infty$, $\operatorname{Ei}(-\rho)\to 0^-$ and
$r(\rho)\to 0^+$, hence the effective static metric has a Killing horizon at
\begin{equation}
r_H=0.
\end{equation}
For a static observer in a metric
$ds^2=-F(r)\,dt^2+F(r)^{-1}dr^2+r^2d\Omega^2$,
the proper hovering acceleration follows, see e.g., Eq.~(19) of Doughty
\cite{1981AmJPh..49..412D}, as
\begin{equation}
a_{\mathrm{hover}}(r)=\frac{|F'(r)|}{2\sqrt{F(r)}}.
\end{equation}
For the metric function $F(r)$ in Eq.~(\ref{eq:inv_Fr}), this gives
\begin{equation}
a_{\mathrm{hover}}(r)=2\left|\operatorname{Ei}^{-1}(-2r)\right|.
\end{equation}
This diverges at the horizon, since
\begin{equation}
a_{\mathrm{hover}}(r)
=
2\left|\operatorname{Ei}^{-1}(-2r)\right|
=
2\rho,
\end{equation}
and therefore
\begin{equation}
a_{\mathrm{hover}}(r_H)=\lim_{\rho\to\infty}2\rho \to \infty,
\end{equation}
similar to the Schwarzschild case. However, unlike Schwarzschild, the surface
gravity of the metric, computed via $\kappa=\frac{1}{2}\left|F'(r_H)\right|,$ vanishes:
\begin{equation}
\kappa
=
\lim_{r\to r_H}\sqrt{F(r)}\,a_{\mathrm{hover}}(r)
=
\lim_{\rho\to\infty} e^{-\rho}(2\rho)
=
0,
\end{equation}
which is more similar to the extremal Reissner--Nordstr\"om (RN) case. Thus, the metric function $F(r)$ in
Eq.~(\ref{eq:inv_Fr}) has a horizon at $r=0$, divergent hovering acceleration,
and vanishing surface gravity.

\begin{center}
\[
\renewcommand{\arraystretch}{1.5}
\begin{array}{c|c|c}
& a_{\rm hover}(r_H) & \textrm{Surface gravity } \kappa \\
\hline
\text{Metric Eq.~(\ref{eq:inv_Fr})} & \infty & 0 \\
\hline
\text{Extremal RN} & \textrm{finite} & 0 \\
\hline
\text{Schwarzschild} & \infty & \textrm{finite}
\end{array}
\]
\end{center}

\begin{figure}[h]
    \centering
    \includegraphics[width=0.45\textwidth]{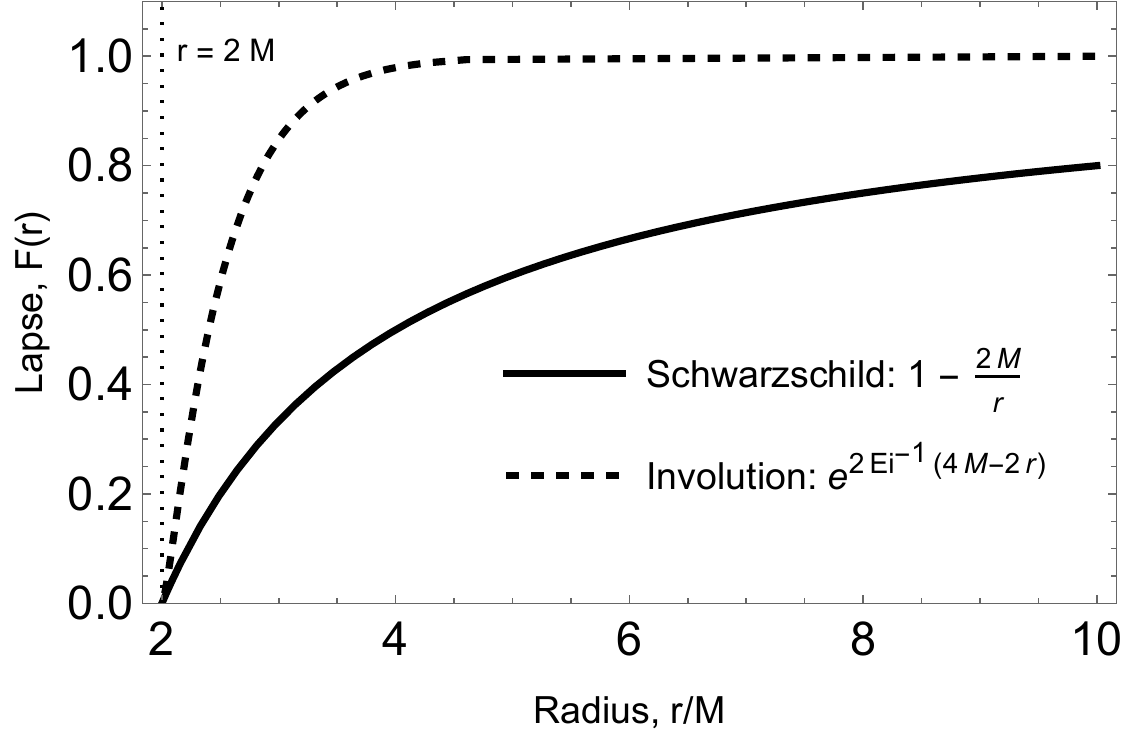}
    \includegraphics[width=0.45\textwidth]{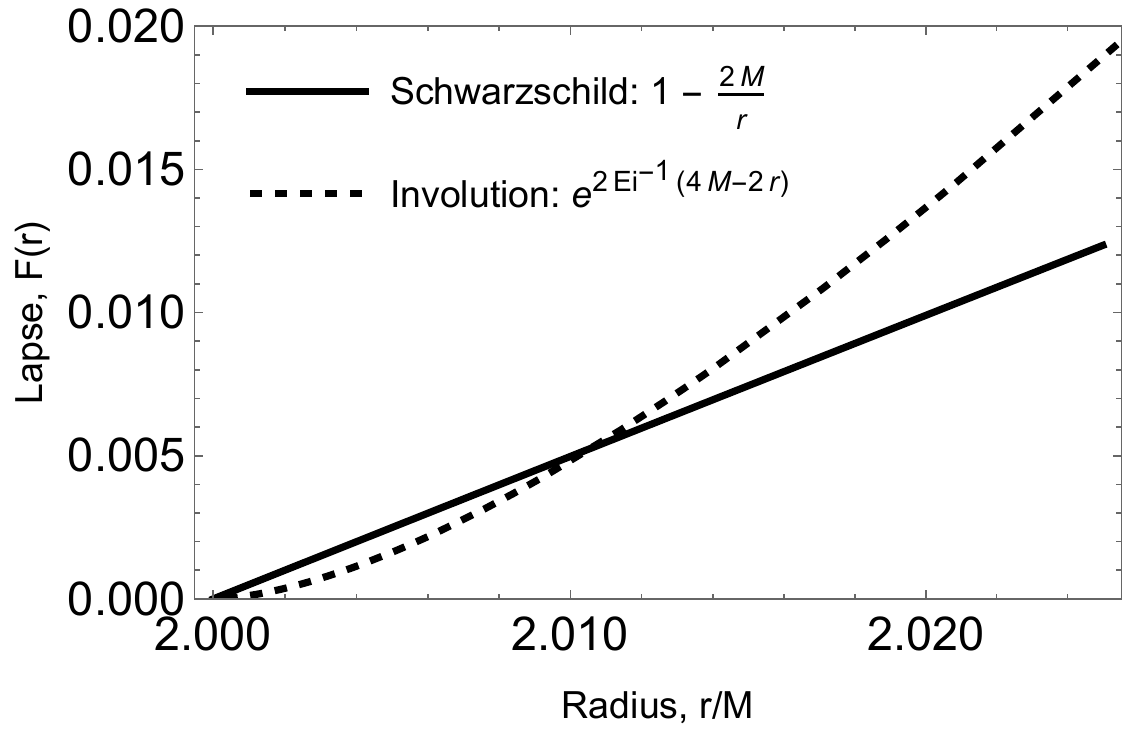}
    \caption{Lapse function of the metric, comparing with the Schwarzschild geometry. The solid curve shows the standard Schwarzschild lapse \(F(r)=1-2M/r\), while the dashed curve shows the lapse function of the new spacetime written in terms of areal radius \(r\). The dashed curve is the shifted lapse function \(F_{\rm new}(r)=\exp\!\big(2\,\Ei^{-1}(v_0-2r)\big)\), where \(v_0\) denotes the advanced-time location of the null shell; here \(v_0=4M\) has been chosen so that the horizon is aligned at \(r_H=2M\), matching Schwarzschild. 
    The bottom panel zooms in near the horizon, showing that the slope of the lapse function (and hence the surface gravity) in our metric approaches zero. 
    \\     } 
    \label{fig:lapse}
\end{figure}

Unlike Schwarzschild or extremal RN, the present construction allows an arbitrary finite-radius black hole horizon; with the choice $v_0 = 0$, the present horizon is compressed to $r_H=0$, so the horizon sphere is reduced to zero area. That choice might naturally be viewed as a type of `pinched' spacetime. The choice of $v_0 = 4M$ places the horizon at $r_H = 2M$, more physically intuitive and in line with the Schwarzschild solution. 
See Fig.~\ref{fig:lapse} for a comparison of the lapse function $F(r)$ to the Schwarzschild case. 

\subsection{Curvature Invariants}
For a static metric
\begin{equation}
ds^2=-F(r)\,dt^2+F(r)^{-1}dr^2+r^2d\Omega^2,
\end{equation}
the curvature invariants are
\begin{align}
R&=-F''-\frac{4F'}{r}+\frac{2(1-F)}{r^2},\\
R_{\mu\nu}R^{\mu\nu}
&=
\frac12\left(F''+\frac{2F'}{r}\right)^2
+
2\left(\frac{F'}{r}+\frac{F-1}{r^2}\right)^2,\\
K&=(F'')^2+\frac{4(F')^2}{r^2}+\frac{4(F-1)^2}{r^4}.
\end{align}
For the involution metric,
\begin{equation}
r(\rho)=-\frac12 \Ei(-\rho),
\qquad
F(\rho)=e^{-2\rho},
\end{equation}
so that
\begin{equation}
F'(r)=4\rho e^{-\rho},
\qquad
F''(r)=8\rho(\rho-1).
\end{equation}
Since \(r\sim e^{-\rho}/(2\rho)\) as \(\rho\to\infty\), one finds
\begin{equation}
R\sim \frac{1}{r^2},
\qquad
R_{\mu\nu}R^{\mu\nu}\sim \frac{1}{r^4},
\qquad
K\sim \frac{1}{r^4}.
\end{equation}
Hence, for the special choice $v_0=0$, then $r=0$ is a genuine curvature singularity. The redshift surface and the singularity are in the same place. That is unusual. 

Unlike Schwarzschild or extremal Reissner--Nordstr\"om, where the horizon is regular, and the singularity lies deeper inside, the present geometry's horizon itself is singular. 

More generally, since the null-shell parameter places the horizon at an arbitrary finite radius $r_H=v_0/2$, the singular-horizon conclusion is not tied to the special gauge choice $v_0=0$: the would-be horizon at $r=r_H$ likewise coincides with a curvature singularity.  Near \(r=r_H\), \(F(r)\sim 4(r-r_H)^2\!\left[\ln\!\left(\frac{1}{r-r_H}\right)\right]^2\), so the tortoise integral \(\int dr/F(r)\) diverges. Therefore, no outgoing null ray emerges from the singular surface in finite exterior time, and the singularity is horizon-bound rather than naked.

\subsection{Anisotropic vacuum-like source interpretation}

If the involution metric is interpreted through Einstein's equation,
\begin{equation}
G^\mu{}_\nu=8\pi G\,T^\mu{}_\nu,
\end{equation}
then the required source takes the anisotropic-fluid form
\begin{equation}
T^\mu{}_\nu=\mathrm{diag}(-\rho_{\mathrm{eff}},\,p_r,\,p_t,\,p_t).
\end{equation}
For a static metric
\begin{equation}
ds^2=-F(r)\,dt^2+F(r)^{-1}dr^2+r^2d\Omega^2,
\end{equation}
the effective density and pressures are
\begin{align}
\rho_{\mathrm{eff}}(r)&=\frac{1-F-rF'}{8\pi G\,r^2},\\
p_r(r)&=\frac{F+rF'-1}{8\pi G\,r^2}=-\rho_{\mathrm{eff}}(r),\\
p_t(r)&=\frac{1}{8\pi G}\left(\frac{F''}{2}+\frac{F'}{r}\right).
\end{align}
Thus the source is radially vacuum-like, with equation of state
\begin{equation}
p_r=-\rho_{\mathrm{eff}},
\end{equation}
but with unequal tangential pressure, so the stress tensor is intrinsically anisotropic. Although the source is anisotropic in the sense that the radial and tangential pressures are unequal, it remains fully spherically symmetric, since the two angular pressures are identical and all quantities depend only on $r$.

Inserting $\ell$, for a scaled involution metric,
\begin{equation}
r(\rho)=-\frac{\ell}{2}\Ei(-\rho),
\qquad
F(\rho)=e^{-2\rho},
\end{equation}
one finds
\begin{equation}
F'(r)=\frac{4\rho e^{-\rho}}{\ell},
\qquad
F''(r)=\frac{8\rho(\rho-1)}{\ell^2},
\end{equation}
and therefore
\begin{equation}
\rho_{\mathrm{eff}}(\rho)=
\frac{1-e^{-2\rho}+2\rho e^{-\rho}\Ei(-\rho)}
{2\pi G\,\ell^2\,\Ei(-\rho)^2},
\end{equation}
\begin{equation}
p_r(\rho)=-\rho_{\mathrm{eff}}(\rho),
\end{equation}
\begin{equation}
p_t(\rho)=
\frac{\rho}{2\pi G\,\ell^2}
\left[
\rho-1-\frac{2e^{-\rho}}{\Ei(-\rho)}
\right].
\end{equation}

The scaling can be understood better near the singular surface \(r\to0^+\). Since
\begin{equation}
r\sim \frac{\ell}{2}\frac{e^{-\rho}}{\rho}
\qquad (\rho\to\infty),
\end{equation}
one obtains
\begin{equation}
\rho_{\mathrm{eff}}(r)\sim \frac{1}{8\pi G\,r^2},
\qquad
p_r(r)\sim -\frac{1}{8\pi G\,r^2},
\end{equation}
whereas the tangential pressure diverges more slowly:
\begin{equation}
p_t(r)\sim
\frac{3}{2\pi G\,\ell^2}
W\!\left(\frac{\ell}{2r}\right)^2,
\end{equation}
where $W$ is the Lambert $W$-function so $p_t$ diverges softer than $1/r^2$.

Hence, the involution geometry is supported by an anisotropic,
radially vacuum-like source whose overall magnitude is set by the
scale $1/(G\ell^2)$. The singularity, therefore, does not resemble
the center of an ordinary isotropic fluid. This motivates the
nonlinear electrodynamics interpretation developed next.

\subsection{Nonlinear electrodynamics interpretation} 

If one interprets this metric through Einstein's equation, the effective stress tensor admits a parametric reconstruction by a
purely electric nonlinear electrodynamics source.

Using signature \((-,+,+,+)\), consider the action
\begin{equation}
S=\int d^4x\,\sqrt{-g}\left[\frac{R}{16\pi G}-\mathcal{L}(\mathcal{F})\right],
\quad
\mathcal{F}\equiv \frac14 F_{\mu\nu}F^{\mu\nu},
\end{equation}
with purely electric field \(F_{tr}={\mathcal E}(r)\), so that 
\begin{equation}
\mathcal{F}=-\frac12 {\mathcal E}(r)^2.
\end{equation}
For nonlinear electrodynamics one has
\begin{align}
\rho_{\mathrm{eff}}
&=-2\mathcal{F}\,\mathcal{L}_{\mathcal{F}}+\mathcal{L},\\
p_r
&=2\mathcal{F}\,\mathcal{L}_{\mathcal{F}}-\mathcal{L}
=-\rho_{\mathrm{eff}},\\
p_t
&=-\mathcal{L},
\end{align}
where \(\mathcal L_{\mathcal F}\equiv \partial \mathcal L/\partial \mathcal F\) and the field equation integrates to
\begin{equation}
r^2\mathcal{L}_{\mathcal{F}}{\mathcal E}=Q,
\end{equation} 
where $Q$ is an integration constant associated with the conserved electric charge. Hence the geometry reconstructs the source parametrically as
\begin{align}
\mathcal{L}(r)&=-p_t(r),\\
\mathcal{F}(r)
&=-\frac{r^4}{2Q^2}
\bigl[\rho_{\mathrm{eff}}(r)+p_t(r)\bigr]^2.
\end{align}
Therefore
\begin{equation}
\mathcal{L}(\rho)=
\frac{\rho}{2\pi G\,\ell^2}
\left(
1-\rho+\frac{2e^{-\rho}}{\Ei(-\rho)}
\right),
\end{equation}
\begin{equation}
{\mathcal E}(\rho)=
\frac{1-e^{-2\rho}+\rho(\rho-1)\Ei(-\rho)^2}{8\pi G\,Q}.
\end{equation} 
Thus, the involution geometry is supported by a purely electric nonlinear
electrodynamics source, with \(\mathcal{L}(\mathcal{F})\) obtained in parametric form.

A notable feature of the reconstructed source is that the electric field remains finite at the singular surface, \({\mathcal E}(\rho\to\infty)\to(8\pi GQ)^{-1}\). The central singularity is not driven by a divergent Coulomb field, but by the fact that the nonlinear electrodynamic response becomes singular: \(\mathcal F\) stays finite while \(\mathcal L(\rho)\) and hence \(p_t=-\mathcal L\) diverge.

Furthermore, the total field energy of the effective electric source,   
\begin{equation}
E_{\rm tot}=4\pi\int_{0}^{\infty}\rho_{\rm eff}(r)\,r^{2}\,dr\ , \end{equation}
vanishes. Explicitly, we find
\begin{equation}
    E_{\rm tot} =\frac{\ell}{4G}\int_{0}^{\infty}\frac{d}{d\rho}\!\left[(1-e^{-2\rho})\,\Ei(-\rho)\right]d\rho = 0.
\end{equation}
Since $(1-e^{-2\rho})\Ei(-\rho)\to 0$ as $\rho\to 0^{+}$ and also as $\rho\to\infty$, the integral reduces to a vanishing boundary term. Thus, although the local electric field and stress tensor are nontrivial, their net energy, integrated over all space, cancels exactly.

\subsection{Proper distance and throat geometry}

The late-time mirror relation, Eq.~(\ref{double-log}),
\be
\tau \sim \ln\ln u
\ee
has a simple geometric counterpart in the metric correspondence.  Consider the static line element
\be
ds^2 = -F(r)\,dt^2 + \frac{dr^2}{F(r)} + r^2 d\Omega^2,
\ee
with parametric form
\be
r(\rho)=r_H-\frac{1}{2}\,\Ei(-\rho),
\qquad
F(\rho)=e^{-2\rho}.
\ee
The proper radial distance on a constant-time slice is
\be
d\sigma = \frac{dr}{\sqrt{F}}.
\ee
Since
\be
\frac{dr}{d\rho}
=
-\frac{1}{2}\frac{e^{-\rho}}{\rho},
\ee
it follows that
\be
d\sigma
=
\frac{1}{\sqrt{F}}\frac{dr}{d\rho}\,d\rho
=
-\frac{1}{2}\frac{d\rho}{\rho}.
\ee
Hence
\be
\sigma = -\frac{1}{2}\ln\rho + \text{const},
\ee
so that the magnitude of the proper distance grows as
\be
|\sigma| \sim \frac{1}{2}\ln\rho.
\ee
Using $\rho=e^\tau$, we obtain
\be
|\sigma| \sim \frac{1}{2}\tau + \text{const}.
\ee
Thus the mirror proper time is, up to a constant factor, equal to the proper radial distance down the associated throat.

This also gives the near-horizon scaling in terms of the radial-coordinate gap.  For large $\rho$,
\be
\Ei(-\rho)\sim -\frac{e^{-\rho}}{\rho},
\ee
so that
\be
r-r_H \sim \frac{1}{2}\frac{e^{-\rho}}{\rho}.
\ee
Equivalently,
\be
\frac{1}{2(r-r_H)} \sim \rho e^\rho.
\ee
Taking the logarithm,
\be
\ln\!\left(\frac{1}{2(r-r_H)}\right)
\sim
\rho+\ln\rho
\sim
\rho,
\ee
where $\ln\rho$ is subleading at large $\rho$.  Therefore
\be
|\sigma| \sim \frac{1}{2}\ln\rho
\sim
\frac{1}{2}\ln\ln\!\left(\frac{1}{2(r-r_H)}\right).
\ee
Thus, the mirror scaling $\tau\sim\ln\ln u$ is reflected geometrically by a double-logarithmic divergence of proper radial distance in the near-horizon coordinate gap.

The main significance of this result is that the mirror’s late-time double-log kinematic scaling becomes a geometric statement: the associated throat is highly compressed in areal radius, so approaching the horizon in $r$ corresponds to only double-log slow growth in proper distance.

\section{Conclusions} 

The question of whether infinite acceleration necessitates infinite energy production has been definitely resolved in the negative by the model presented here, completing the third case of the triangle presented in the Introduction. In analyzing this infinite acceleration trajectory, we find it 
provides an analytically tractable example of radiation from extreme acceleration with several interesting features:
\begin{itemize}
\item \textbf{Exponential rapidity growth:} The mirror’s rapidity grows exponentially with proper time,
$\eta(\tau) = e^\tau$, following from the dynamical condition $\alpha(\tau) = \eta(\tau)$.

\item \textbf{Symmetric ray scattering:} The null ray tracing functions satisfy $p = f^{-1}$, and $u\leftrightarrow v$, so the scattering map is an involution with $p(p(u)) = u$.

\item \textbf{Finite energy emission:} Despite unbounded acceleration, the mirror emits a finite total energy $E = 1/(12\pi)$.

\item \textbf{Radiation localized near Doppler matching:} The dominant contribution to radiation occurs when the rapidity matches the logarithmic frequency ratio, $\rho \approx \lambda$. 

\item \textbf{Agreement between stress tensor and particle summation:} The total emitted energy obtained from the flux exactly matches the quantum summation.

\end{itemize}
This mirror is a rare and relatively tractable model where involutive null scattering, extreme acceleration, finite total radiation, horizon formation, and the distinction between local flux and global particle weighting can all be seen in one place. 

We have presented and studied the behavior of the Bogoliubov coefficients for particle production. We also presented the spacetime metric associated with the accelerating trajectory, and discussed the singularity structure. This model presents an intriguing mix of normal and extremal black hole properties, and a double exponential (or double log, depending on your point of view) warping of the spacetime proper time vs.\ radial infall relation.

\section{Acknowledgements} 
\noindent Funding of M.G. comes partly from the FY2024-SGP-1-STMM Faculty Development Competitive Research Grant (FDCRGP) no.201223FD8824 and SSH20224004 at Nazarbayev University in Qazaqstan.  M.G. gives appreciation to the ROC (Taiwan) Ministry of Science and Technology (MOST), Grant no.112-2112-M-002-013, National Center for Theoretical Sciences (NCTS), and Leung Center for Cosmology and Particle Astrophysics (LeCosPA) of National Taiwan University.

\clearpage
\bibliography{main} 

\end{document}